\documentclass[twocolumn,pre,superscriptaddress]{revtex4}
\usepackage[paperwidth=210mm,paperheight=297mm,hmargin=2cm,vmargin=2.5cm]{geometry}

\usepackage{graphicx}
\usepackage{epstopdf}
\usepackage{amsmath}
\usepackage[english]{babel}
\usepackage[justification=raggedright,singlelinecheck=false]{subfig}

\begin{document}
\title{Morphological Control of Grafted Polymer Films by Nanoparticle Binding}
\author{Michael G. Opferman}
\affiliation{Department of Physics and Astronomy, University of Pittsburgh, Pittsburgh, Pennsylvania 15260, USA}
\author{Rob D. Coalson}
\affiliation{Department of Chemistry, University of Pittsburgh, Pittsburgh, Pennsylvania 15260, USA}
\author{David Jasnow}
\affiliation{Department of Physics and Astronomy, University of Pittsburgh, Pittsburgh, Pennsylvania 15260, USA}
\author{Anton Zilman}
\affiliation{Theoretical Division, Los Alamos National Laboratory. POB 1663, Los Alamos, New Mexico 87545, USA}
\affiliation{Department of Physics, University of Toronto, Toronto, Ontario M5S 1A7, Canada \footnote{Current Address}}

\date{\today}

\begin{abstract}
Mixtures of nanoparticles and polymer-like objects are encountered in many nanotechnological applications and biological systems. We study the behavior of grafted polymer layers decorated by nanoparticles that are attracted to the polymers using lattice gas based mean field theory and accompanying coarse-grained Brownian dynamics simulations. We find that the presence of nanoparticles can induce large morphological transitions in the layer morphology. In particular, at moderate nanoparticle concentrations, the nanoparticles cause a reduction in the height of the polymer layer above the grafting surface, which occurs via a novel first-order phase transition for sufficiently strong attraction between the polymers and the nanoparticles and  smoothly for weak attractions. The predictions of the theory qualitatively agree with the observed behavior of grafted natively unfolded protein strands upon binding of proteins \cite{Lim2007}. The results also inform ways of designing nanopolymer layer morphologies.

\end{abstract}
\maketitle

\section{Introduction}
Polymeric materials play an important role in materials science and engineering.  Along with traditional polymeric materials such as rubbers and plastics, thin films of polymers, often grafted to surfaces to form ``polymer brushes," have become prominent in many technological applications such as those reviewed by Senaratne in Ref.$\;$\cite{BrushBiotechReview}.  Applications include stabilizing a disperse colloidal suspension \cite{ColloidReview,ColloidalSuspensions} or creating devices like novel electrodes \cite{ColloidElectrodes}, organic solar cells \cite{SolarCells}, or organic memory \cite{PolymerMemory}.  Overall, the properties of such thin polymer films have been well characterized experimentally and reasonably well understood theoretically (e.g. \cite{deGennes1980, MilnerWittenCates, Murat1989, Milner1990, Milner1991, Kim2006, Binder2009}).

Thin polymer films decorated with nanoparticles are a novel class of composite materials which allow one to design their function and control their morphology in nanotechnological applications such as solar cells and programmable memory devices \cite{NanoparticleComposites,NanoparticleMemory,ColloidElectrodes,SolarCells}.   A class of such additives, sometimes called ``sticky" antiplasticizers, which bind to polymers, has been studied experimentally (e.g. \cite{Narkis1996,Monnerie1998,Halary2002}). These additives form hydrogen bonds with the polymers to which they are added and thus decrease the free volume in the material \cite{Narkis1996}.  Their addition results in modification of the crosslinking in the polymer network and changes in the cooperative motion between polymers \cite{Monnerie1998}.  Practically speaking, they modify the crystallinity, glass transition temperature, toughness, and elastic moduli of polymeric materials.  Only recently has the addition of such additives to a polymer melt been studied in molecular simulations \cite{Pablo2007, Dasmahapatra2009}.

Polymer-like biological molecules, such as natively unfolded proteins, play an important role in the functioning of living cells.  There are many examples, such as those reviewed by Uversky in Ref.$\;$\cite{UnfoldedReview}.  Examples include newly-made proteins  in the endoplasmic reticulum \cite{UnfoldedER}, a protein implicated in Alzheimer's Disease called NACP \cite{NACP}, and nucleoporins which facilitate transport through the nuclear pore complex (NPC) \cite{Macara2001, Kutay2003, Rexach2007, Stewart2007}.  In the case of nucleoporins, it is common for these biopolymers to bind to other proteins \cite{Stewart2005, Schulten2005}, which we think of as analogous to nanoparticles.  For instance, it has been hypothesized \cite{Lim2007, Lim2006, Lim2008} that transport through the NPC involves a conformational change in the layer of natively unfolded proteins that line the NPC passageway, induced by the binding of certain transport factor proteins. Such changes have been observed in \textit{in vitro} experiments \cite{Lim2007}.  However, their precise role in the transport of cargoes through the NPC is still not understood.

Without specific attractive interactions between the nanoparticles and the polymers, they effectively repel each other, and the effect of this repulsion on nanoparticle aggregation is an important element in controlling the physical properties of polymeric materials (see e.g. \cite{NanoparticleComposites} and references therein).  Kim and O'Shaughnessy \cite{Kim2006} derived the spatial distribution of such nanoinclusions in a polymer brush $-$ for instance whether they interpenetrate the brush or segregate into a layer on top of the polymers.  Similar ideas were examined in computer simulations by Binder and colleagues \cite{Binder2009}.

The case in which nanoparticles are attracted to grafted polymers has received less attention.  Marko considered a polymer brush in the presence of a binary mixture of solvents \cite{Marko1993}, which could, in principle, model this system.  However, he focused primarily on the case of repulsive interactions and the case in which attractive interactions occur via higher order terms in the density expansion as compared with the case considered here.  More recently, in Ref.$\;$\cite{Halperin2011}, Halperin \textit{et al.} considered, among other things, nanoparticles that bind weakly to a polymer brush.  The weak binding assumption results in a behavior devoid of the many features described in our work, such as global morphology changes.

	The main goal of the present study is to gain insight into the behavior of polymers grafted to a flat surface interacting with nanoparticles which can bind to the brush. We consider a situation in which the polymer layer is in equilibrium with a solution of nanoparticles which can penetrate into the layer, thereby changing properties such as the height of the layer above the grafting surface and its density. Our goal is to establish how the properties of the polymeric layer, particularly its height above the grafting surface, depend on the number of nanoparticles bound inside the layer, the strength of their interaction with the polymers, and ultimately the concentration of the particles in the solution above the layer.  We will model the system using a lattice gas based mean field theory and use this theory to describe the collapse of the polymer chains upon binding.  We also will present Brownian dynamics simulations, which were performed in order to validate and further elucidate the behavior predicted by the mean field theory.

\section{Mean Field Theory} \label{MFT}
In this section we present a general theory of the behavior of polymer layers mixed with nanoparticles.  We will first obtain a free energy for the system using an approach based on a lattice gas model \cite{HillBook}, and then we will examine in detail the features of this free energy and the equilibrium states predicted by it.
\subsection{Free Energy} \label{FreeEnergy}
Consider a layer of $N_p$ polymeric chains, each containing $N$ monomers, which are grafted at one end to a planar surface. In the absence of direct interactions (beyond excluded volume) between the monomers of the chains, such planar layers are known as ``brushes'' and have been extensively studied (e.g. \cite{deGennes1980, MilnerWittenCates, Murat1989, Milner1990, Milner1991, Kim2006, Binder2009}).

Our initial goal is to determine the free energy of the polymer layer
as a function of the essential variable quantities, namely the layer's height $h$ and the number of nanoparticles $N_\ell$ in the layer volume $M_\ell$. We will obtain the entropy by first
considering a different but related system comprised of $N_m=N_pN$ disconnected monomers and $N_\ell$ nanoparticles in a volume of $M_\ell$
elemental cells (each with volume $b^3$ where $b$ is the monomer diameter), and then correcting it for the effect of polymer
connectivity (including grafting).  One could also obtain a similar expression for the entropy using other methods (as discussed below).

Here, we consider the case in which the nanoparticles have the same size as monomers.  We have also considered the more general case in which nanoparticles are larger than monomers \cite{fatkaps}. We find that the mean field theory predicts the same qualitative features, including the rich behavior of subsections \ref{Landscape} and \ref{collapse}.  We therefore focus on the simpler case for the sake of clarity in the present work.

In a simple lattice gas model for a system of $N_\ell$ nanoparticles and $N_m$ disconnected monomers of the same size, the total number of configurations is
\begin{equation} \label{multinomial}
\Omega=\frac{M_\ell!}{N_\ell!N_m!(M_\ell-N_\ell-N_m)!}.
\end{equation}
The entropic contribution to the free energy, in units of $kT$, is therefore given by $F_{ent}=-\ln\Omega$.  Making use of Sterling's approximation and converting to intensive variables,
\begin{eqnarray}\label{Fent1_c2} \nonumber
  F_{ent} &=& M_\ell \{ \psi \ln \psi +  \phi \ln \phi \\
&&+ (1-\phi - \psi) \ln[1 - \phi - \psi ] \}
\end{eqnarray}
where $\psi = N_m/M_\ell$ is the density (or, since we express all lengths in units of $b$, the volume fraction) of monomers in the polymer layer and $\phi = N_\ell/M_\ell$ is the density (or volume fraction) of the nanoparticles.  This expression includes contributions from the translational entropy of monomers and nanoparticles, as well as excluded volume in that no two particles may occupy the same lattice site.

However, the monomers are not a gas, but rather are linked together into  polymer chains, which, in addition, are tethered to a flat surface.  To correct the above
results for polymer connectivity and grafting \cite{deGennes1980, RussianBook}, the translational entropy of the monomers should be replaced by the expression for the
configurational entropy of an ideal polymer chain grafted onto a flat
surface as in a polymer brush, namely $ -N_p h^2 / (2 N)$ \cite{MilnerWittenCates}. Strictly speaking, the translational entropy of the monomers is $-M_\ell \psi (\ln \psi -1)$, but since adding a constant to the free energy will not change the equilibrium states, we will, for simplicity, treat $-M_\ell\psi \ln \psi$ as the translational entropy of an ideal gas of monomers.  For conveniece, we will henceforth implicitly normalize free energies by the number of chains $N_p$.  The entropic contribution to the layer free energy per polymer chain (in units of $kT$ and denoting $M=M_\ell/N_p$) then becomes
\begin{eqnarray} \label{Fent3} \nonumber
  F_{ent}  &=& h^2 / (2 N) + M \{ \phi \ln \phi   \\
&& + (1-\phi - \psi) \ln[1 - \phi - \psi ] \}.
\end{eqnarray}

It should be noted that our lattice gas based derivation is not the only way to obtain the entropic part of the free energy of the system.  A derivation based on Flory's approach \cite{Flory1942,FloryBook}, in which monomers are placed on adjacent lattice sites in accordance with polymer connectivity, would give a free energy substantially the same as Eq.$\;$(\ref{Fent3}), with only inessential differences such as terms that do not contribute to its relevant intensive thermodynamic quantities and terms which disappear in the limit $N\gg1$.  A van der Waals gas based approach for a multicomponent system \cite{OnukiBook}, which does not rely on a lattice, can also be used.  The van der Waals analog of Eq.$\;$(\ref{Fent3}) would be slightly different in appearance, but would have substantially the same virial expansion to second order in concentration of monomers and nanoparticles as well as similar behavior at high concentration.

Also note that when $\phi =0$ in Eq.$\;$(\ref{Fent3}) (no nanoparticles in the polymer layer)
\begin{equation}\label{Fpoly}
F_{ent}
\rightarrow h^2 / (2 N) + M(1-\psi)\ln[1-\psi].
\end{equation}
If the logarithm is expanded for $\psi\ll1$, this becomes the standard mean field free energy of a plane-grafted polymer \cite{MilnerWittenCates, Milner1991}, and it yields the standard brush height scaling behavior $h \sim Na^{-2/3}$  \cite{deGennes1980, MilnerWittenCates, Milner1991}.  We choose not to expand the logarithm because the density may become large once nanoparticles are added.
The first term in Eq.$\;$(\ref{Fpoly}) describes the entropic cost of
stretching the grafted polymer segment. The second term accounts
for excluded volume between monomers, with a built-in upper limit to the
density $\psi = 1$.
(At this density monomers fill up the entire lattice, and thus the
polymer layer cannot be compressed any further.)

To include the contribution to the free energy due to the
attraction between monomers and nanoparticles, we
will write for the net interaction energy between one nanoparticle
and the polymers
$\chi(\epsilon_b)\psi$, where $\chi$ is a phenomenological interaction constant which depends  on the strength of the microscopic interaction between the monomers $\epsilon_b$ (discussed in subsection \ref{SimCollapse} below).
Thus, if there are $N_\ell=M_\ell\phi$ nanoparticles in the polymer layer, the total
internal energy of all such interactions is, in units of $kT$ and normalized by the number of chains,
\begin{equation}\label{Fkap-nup}
\bar{E} =  M \chi \phi\psi.
\end{equation}
Note that $\chi <0$ corresponds to attractive
interactions. Finally, the total
free energy of the polymer brush layer is $F_B = \bar{E}  +
F_{ent}$, with $F_{ent}$ given in Eq.$\;$(\ref{Fent3}), i.e.,
\begin{eqnarray} \label{FB} \nonumber
  F_B  &=&   h^2 / (2 N) + M \{ \chi \phi \psi + \phi \ln \phi   \\
&& + (1-\phi - \psi) \ln[1 - \phi - \psi]  \}.
\end{eqnarray}

On one side of the polymer layer is the grafting plane.  On the other side (``above" the layer) is a solution of nanoparticles.
In addition to the free energy of the polymer layer, we also need to include the
free energy of the solution that is in equilibrium with the polymer layer.  We will
model the solution as an ideal solution comprised of
nanoparticles mixed with an unstructured
solvent. The free energy of the solution, $F_S$, is standard, and is given by the free energy of mixing of a gas of nanoparticles, and
hence by the r.h.s. of Eq.$\;$(\ref{Fent1_c2}) with $\psi=0$.

Denoting
$M_S$ as the volume of the solution (i.e., the number of
monomer sized cells that comprise the solution volume), and
$N_S$ as the number of nanoparticles in the solution, we
compute for the solution chemical potential
\begin{equation}\label{mu-soltn}
\mu_S = \partial F_S (M_S,N_S) / \partial N_S  = \ln \left[ C_0
/(1 -  C_0 ) \right],
\end{equation}
where $C_0 =  N_S / M_S$ is the
density (or, equivalently in our units, the volume fraction) of nanoparticles in the solution. Similarly, the osmotic pressure of nanoparticles in
the solution is given by
\begin{equation}\label{p-soltn}
\Pi_S = -\partial F_{S} (M_S,N_{S}) / \partial M_{S} =
- \ln \left[ 1 - C_0 \right].
\end{equation}

In the same way, the chemical potential and osmotic pressure in the polymer layer can be obtained from Eq.$\;$(\ref{FB}) as
\begin{equation}
\label{mu B}
\mu_{B}= \chi \psi+ \ln[\phi/(1-\phi - \psi)]
\end{equation}
\begin{equation}
\label{Pi B}
\Pi_B = \chi \phi \psi - \frac{ h}{a^2N} - \psi - \ln(1-\phi - \psi).
\end{equation}

In thermodynamic equilibrium, the chemical potential of the nanoparticles and the osmotic pressure must
be equal in the polymer layer and in solution, which are assumed to be macroscopic.  Consequently, the chemical potential of the nanoparticles in the layer and osmotic
pressure of the polymer layer are determined by the bulk solution
concentration $C_0$, as indicated in Eqs.$\;$(\ref{mu-soltn}) and
(\ref{p-soltn}) above.  Equating the chemical potentials and osmotic pressures of the two phases
gives two equations in the
two unknown variables $M$ and $N_B$ (where $N_B=N_\ell/N_p$).  Simultaneously solving these two equations yields the polymer layer's volume and number of bound nanoparticles at the extrema of the free energy.

One can ensure that the solution found is the global minimum of the free energy by instead minimizing the total thermodynamic potential
\begin{eqnarray}\label{F-tot}\nonumber
\Phi (M,N_B) &=& F_B (M,N_B) - \mu_S (C_0) N_B  \\
&& + \Pi_S (C_0) M
\end{eqnarray}
over $M$ and $N_B$ with $C_0$ held fixed. This is analogous to minimizing the grand potential $F-\mu N$, but allows for variation of the polymer layer's volume as well.
By direct differentiation it can be verified
that the extrema of $\Phi$ correspond to equal chemical potential and equal osmotic pressure in the layer and in solution.

\subsection{Free Energy Landscape}\label{Landscape}
Setting $\mu_B$ and $\mu_S$ equal yields an explicit solution for $\phi$.  This gives, in equilibrium,
\begin{equation}
\label{eq condition}
\phi = \frac{(1 - \psi)\gamma}{\gamma+\exp(\chi \psi)}
\end{equation}
with $\gamma = C_0/(1-C_0)$.

If one makes use of Eq.$\;$(\ref{eq condition}) to reduce $\Phi(M,N_B)$ to one variable, one obtains
\begin{eqnarray} \label{Ftrans1D} \nonumber
  \Phi  &=&   h^2 / (2 N) + M (1-\psi)\ln[1-\psi] \\
&&-M\ln[1-C_0] \\
\nonumber
&& - M(1-\psi)\ln[1+\gamma \exp(-\chi \psi)].
\end{eqnarray}
Note that the first two terms are simply the free energy obtained for a polymer brush with no nanoparticles, Eq.$\;$(\ref{Fpoly}).  Since $\psi=N/M$, this form allows $\Phi$ to be regarded as a function of one variable, $M$ (or, equivalently, $h=M/a^2$ where $a$ is the distance between grafting sites), with $C_0$ as a fixed input parameter.

\begin{figure}[h]
\subfloat{
\label{lowchi}
\includegraphics[scale=0.5]{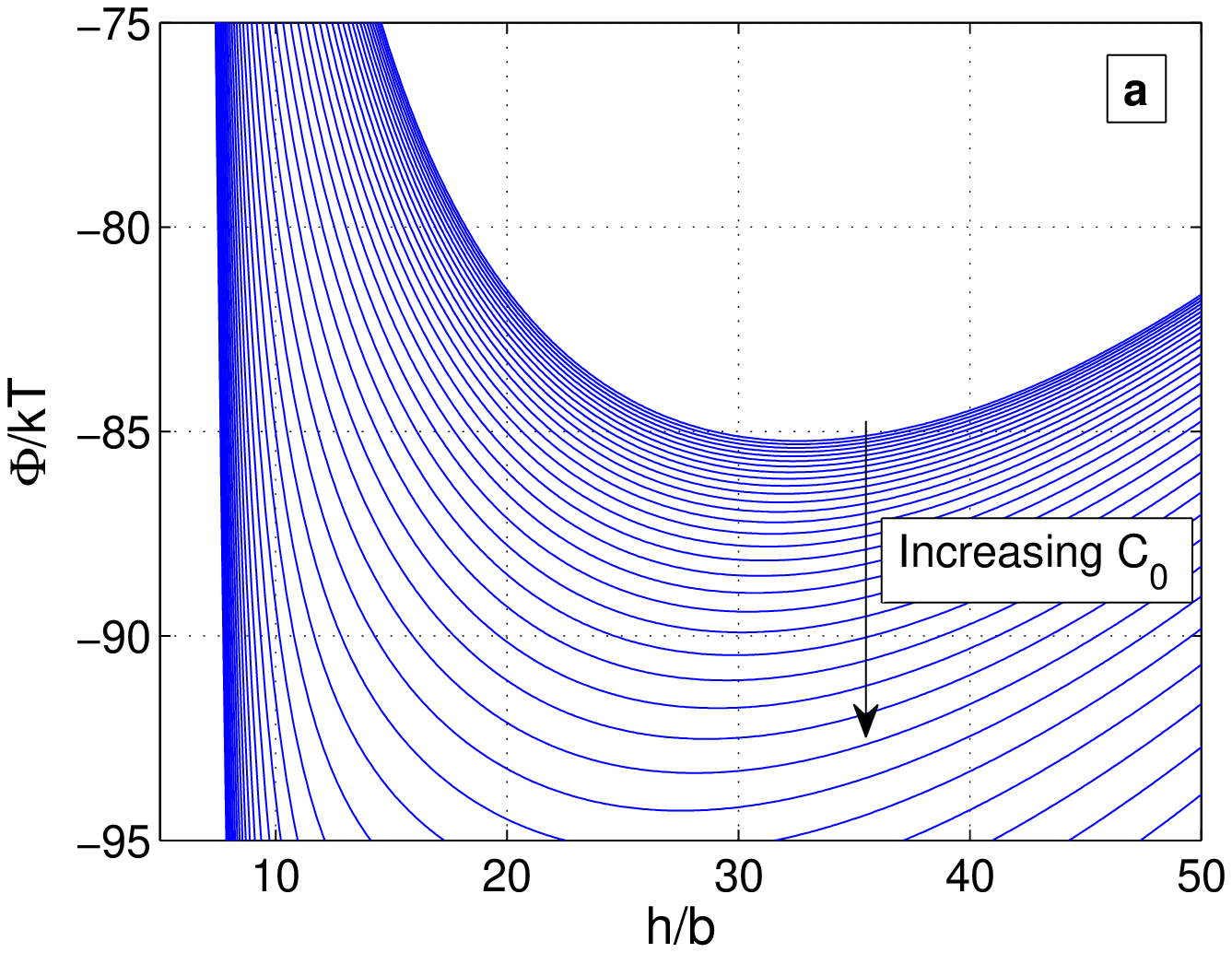}}\\
\subfloat{
\label{highchi}
\includegraphics[scale=0.5]{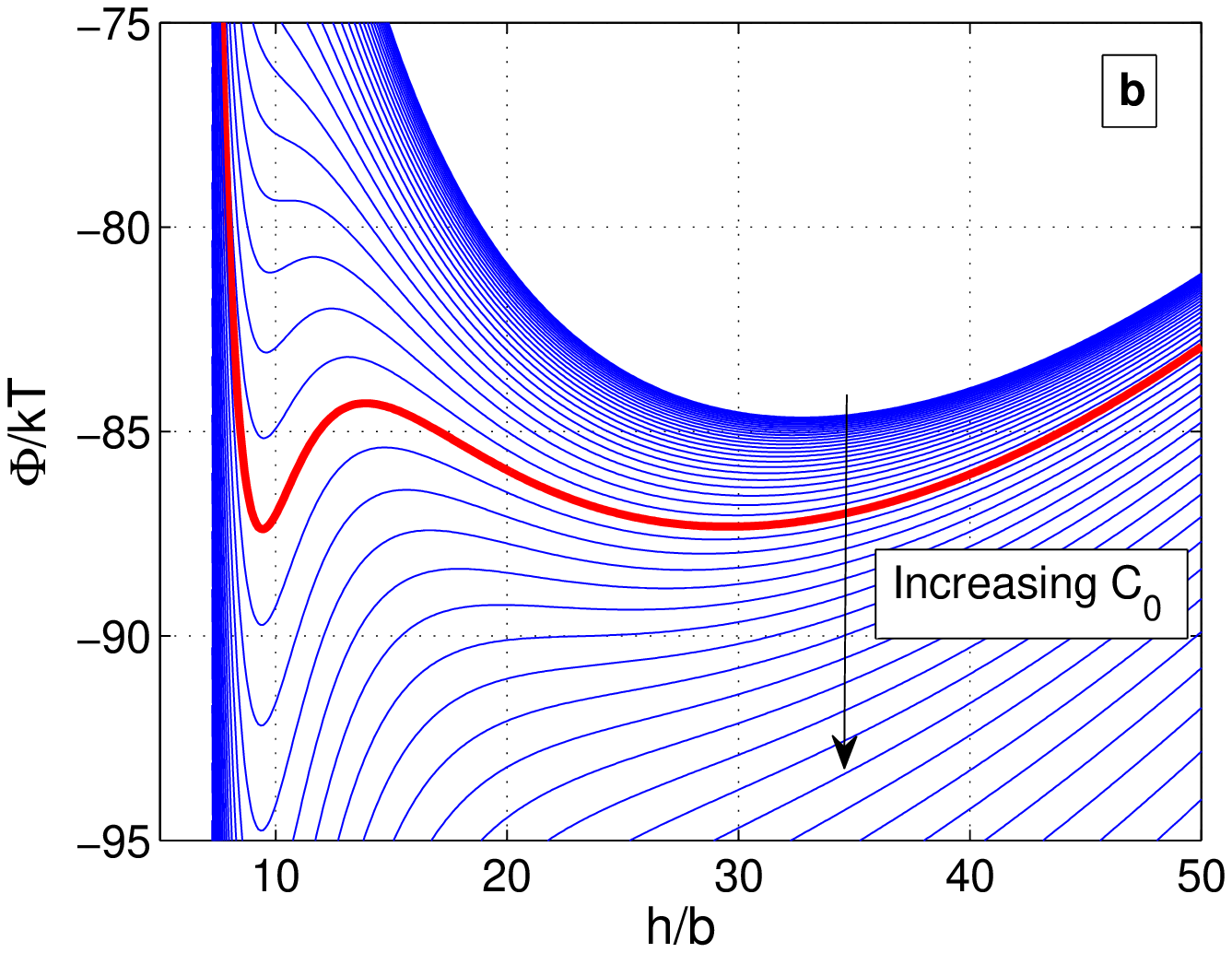}}
\caption{The total thermodynamic potential $\Phi$ vs. polymer layer height $h$ for $a=4$, $N=100$, and (a) $\chi=-6$ and $10^{-3}<C_0<10^{-1}$ or (b) $\chi=-10$ and $10^{-4}<C_0<10^{-2}$.  In (a) there is a single minimum for all $C_0$.  In (b) there exists a range of $C_0$ for which the potential has a double-minimum structure.  At approximately $C_0=1.2\times10^{-3}$ (shown via the bold red line), the two minima are equal in energy (i.e. a phase transition occurs).}
\label{fig2}
\end{figure}

It is instructive to plot $\Phi$ vs. $h$ at fixed $C_0$ as seen in Fig. \ref{fig2} for $\chi=-6$ and $\chi=-10$ (in units of $kTb^3$).  For weak binding strength (e.g. $\chi=-6$ as seen in Fig. \ref{lowchi}), $\Phi$ vs. $h$ for fixed $C_0$ has a single minimum for all $C_0$.  As a result, the height of the polymer layer changes smoothly upon the addition of more nanoparticles.  On the other hand, if the binding strength $|\chi|$ between the nanoparticles and the chains exceeds a certain threshold, the free energy can have two local minima as seen in Fig. \ref{highchi}, with the global minimum representing the equilibrium state.  Within mean field theory, this results in a discontinuous transition in the polymer layer between a high $h$ (extended) and low $h$ (collapsed) state when sweeping through $C_0$.

\subsection{Behavior of the polymer layer's height as a function of particle concentration.} \label{collapse}
\begin{figure}[h]
\includegraphics[scale=0.5]{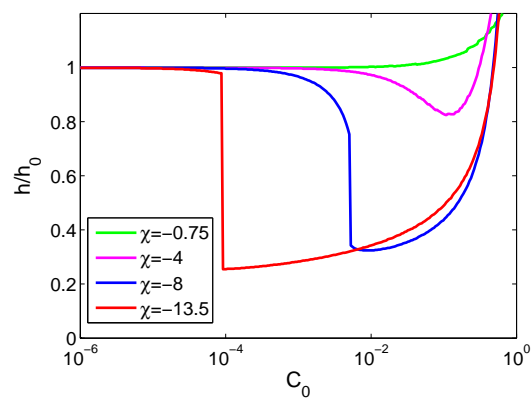}
\caption{Layer height, normalized to its value in the absence of nanoparticles $h_0$, as a function of the nanoparticle concentration in the bulk solution for several values of binding strength $|\chi|$ with grafting distance $a=4$ and $N=100$.}
\label{fig1}
\end{figure}

In this subsection we show the predictions of the mean field theory of subsection \ref{FreeEnergy} regarding the effect of the nanoparticles on the extension of the polymer brush layer at equilibrium.  At moderate nanoparticle concentrations, the nanoparticles typically cause the brush to attain a more compact state.  This decrease in $h$ may occur via a smooth decrease (for weak binding) or a first-order phase transition (for strong binding).  Fig. \ref{fig1}  illustrates the effect of nanoparticles on the height of the polymer layer for various binding strengths.

For sufficiently weak binding, such as $\chi=-0.75$, the effect of nanoparticle addition is only noticeable at very high nanoparticle concentrations.  The rise in layer height at high concentration occurs because the brush must swell in order to accommodate the large number of bound nanoparticles.

For moderate binding strength, such as $\chi=-4$, there is a range of $C_0$ (approximately $10^{-2}<C_0<10^{-1}$) in which the height of the polymer layer decreases smoothly as the nanopartcle concentration increases.  Physically, if the brush is in a more compact state, each nanoparticle will interact, on average, with more monomers.  These attractive interactions help to overcome the excluded volume repulsion between the polymers and thus favor the compact state.

For sufficiently strong binding, such as $\chi=-8, -13.5$, this decrease in height occurs via a sharp transition, as illustrated in Fig. \ref{highchi}.

 Note that a decrease in the height of the polymer layer upon addition of binding particles is qualitatively consistent with the experiment of Ref.$\;$\cite{Lim2007}, where addition of the binding proteins causes a brush of biopolymers to strongly compress.

It might be thought that the nanoparticles simply play the role of solvent molecules, and their presence is analogous to degrading the solvent quality.  However, in a standard mean field analysis of a polymer brush in a variable solvent, one would start with the standard brush free energy in Eq.$\;$(\ref{Fpoly}) and add a term $M \tilde{\chi} \psi^2$, where $\tilde{\chi}$ is the interaction parameter, to account for solvent quality.  Such a free energy does yield a compression of the brush for poor solvent quality ($\tilde{\chi}<0$), but does not contain the phase transition behavior predicted here.
\section{Simulations}\label{Sim}
In order to augment and verify the mean field theory results, we now turn to coarse-grained Brownian dynamics simulations (e.g. \cite{McCammon1978,Shaqfeh2002}) of the system.  In this type of simulation, a particle performs overdamped  biased diffusion under the influence of the deterministic force arising from potential interactions with other particles and the random force that represents the thermal fluctuations due to random collisions with the solvent molecules.
In this paper, we are only interested in equilibrium properties, so that the system, comprised of monomers of the polymer chains and nanoparticles, samples the Boltzmann distribution of configurations at the prescribed temperature. Hence the precise value of the friction constant is not critical, and we can assume sufficiently high friction so that an overdamped Brownian (Langevin) algorithm is appropriate.
\subsection{Interactions} \label{interactions}
	We modeled the polymer strands as chains of beads with nearest-neighbor beads connected by finitely extensible, nonlinear, elastic (FENE) springs \cite{FENE}, which exert a force of the form
\begin{equation} \label{FENE}
F_{FENE}=\frac{-kr}{1-\left(\frac{r}{\ell_{max}}\right)^2}
\end{equation}
where $k$ is the spring constant, $r$ is the distance between beads, and $\ell_{max}$ is the maximum separation of beads, at which the FENE restoring force becomes infinitely strong.  This force behaves approximately as a Hooke's Law spring for $r \ll \ell_{max}$, but becomes very large when $r$ approaches $\ell_{max}$.  As a result, it may be used when spring-like forces are expected to produce the correct physics, but large separations are physically unrealistic.

In addition, all beads interact through a pairwise truncated 6-12 Lennard-Jones (LJ) potential \cite{Leach} that represents short-range repulsion (excluded volume).  We cut off the LJ potential at its minimum so that it is purely repulsive.  Mathematically, the potential has the form
\begin{equation}
\label{cutoff}
U_{trun}=\left\{
\begin{array}{lr}
\epsilon\left( \left( \frac{b}{r}\right)^{12} - 2\left(\frac{b}{r}\right)^6 \right)+\epsilon,\; \; r<b \\
0,\; \; \; \; \; \; \; \; \; \; \; \; \; \; \; \; \; \; \; \; \; \; \; \; \; \; \; \; \; \; \; \; \; \; \;   r>b \end{array} \right.
\end{equation}
In this potential, $\epsilon$ sets the strength of the repulsive force, and $b$ sets its spatial range (and thus the bead size).  Since this LJ potential sets the monomer size, $b$ is taken to be the same as the unit of length in the mean field theory.

	Nanoparticles are modeled as single beads of the same size as polymer beads.  They, too, interact repulsively with each other via Eq.$\;$(\ref{cutoff}), but they feel no analog of Eq.$\;$(\ref{FENE}) and therefore do not form chains.

We consider the case in which nanoparticles and polymer beads interact with each other attractively.  This is achieved via a LJ-type potential which, in addition to the repulsive part, has an attractive part composed of another LJ potential with a different value of $\epsilon$.  The two parts are pieced together continuously at their common minimum ($r=b$).  That is,
\begin{equation}\label{Upn}
U_{pn}=\left\{
\begin{array}{lr}
\epsilon\left( \left( \frac{b}{r}\right)^{12} - 2\left(\frac{b}{r}\right)^6 \right)+\epsilon-\epsilon_b, \; \; r<b \\
\epsilon_b\left( \left( \frac{b}{r}\right)^{12} - 2\left(\frac{b}{r}\right)^6 \right),\; \; \; \; \; \; \; \; \; \; \; \; \; r>b \end{array} \right.
\end{equation}
where $\epsilon_b$ is the binding energy of the attractive nanoparticle-monomer interaction.  The potential in Eq.$\;$(\ref{Upn}) ensures that the strength of the attractive interaction can be varied without simultaneously changing the excluded volume interaction.

	Nanoparticles and monomers both interact with the grafting surface via another truncated, purely repulsive LJ-type potential centered at $z=0$ and truncated at $z=b$.  This force depends on the $z$ coordinate of the particles only, and so models a flat wall.  The positions of the terminal beads of all chains are fixed at $z=0$, so that each chain is rigidly grafted to the surface at one end.

The results reported here are for chains of length $N=100$ beads that were grafted onto a square lattice with lattice parameter $a=4$ or $a=3$, giving a grafting density of $\sigma=1/a^2$.  Periodic boundary conditions were used in the $x$ and $y$ directions.  Unless otherwise noted, a system size of 16 chains ($4 \times 4$) was used.
\subsection{Simulation Measurements and Procedures}
	The simulation box contains the grafted polymer layer, above which there is a solution of nanoparticles free of polymers.  The upper boundary (``ceiling") of the simulation box is placed in contact with a reservoir of nanoparticles.  The solution of nanoparticles above the brush is held at fixed concentration (within fluctuations) by fixing the incoming flux of nanoparticles from the reservoir, while the outgoing flux is determined by making the ``ceiling'' an absorbing boundary. This means that the entire simulation box comes to equilibrium with a reservoir of nanoparticles at a fixed chemical potential (i.e. a grand canonical ensemble is simulated).  For fixed flux into the box, it was verified that the time-averaged concentration remained constant after equilibrium was reached, and this concentration was found to be independent of the properties of the brush (e.g. $h$, $\epsilon_b$).

	The brush height was measured by creating histograms of the monomer density as a function of $z$ such as those in subsection \ref{Profiles}, and marking the top of the brush as the $z$ value at which the monomer density became negligible.
	Once $h$ is determined from the monomer density profile, all nanoparticles with $z<h$ are considered bound, and those with $z>h$ are considered to be in solution.

\section{Simulation Results}\label{SimResults}
\subsection{Behavior of the polymer layer height as a function of particle concentration}\label{SimCollapse}
\begin{figure}[h]
\includegraphics[scale=0.5]{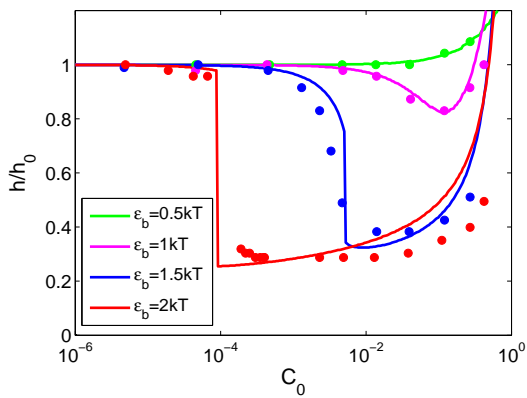}
\caption{$h/h_0$ vs. $C_0$ for four different binding strengths and grafting distance $a=4$.  Grand canonical simulation results (circles) and MFT results (lines) plotted on the same axes. The $\chi$ values for the analytical lines are the same as those of Fig. \ref{fig1} and correspond to the $\epsilon_b$ values shown in the legend.}
\label{overlay4}
\end{figure}

\begin{figure}[h]
\includegraphics[scale=0.5]{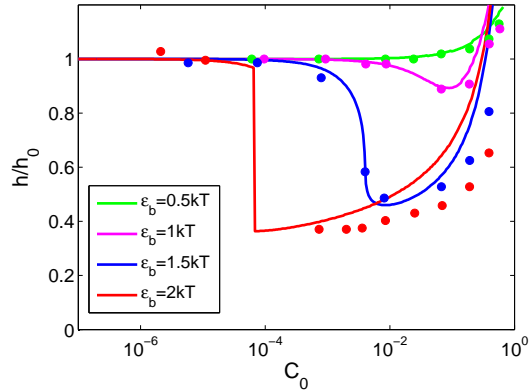}
\caption{$h/h_0$ vs. $C_0$ for four different binding strengths and grafting distance $a=3$. Grand canonical simulation results (circles) and MFT results (lines) plotted on the same axes. The $\chi$ values for the analytical lines are the same as those of Fig. \ref{fig1} and correspond to the $\epsilon_b$ values shown in the legend.}
\label{overlay3}
\end{figure}

Simulation results support the mean field picture, which was illustrated in Fig. \ref{fig1}.   Using the same grafting density as in Fig. \ref{fig1} as well as a second grafting density, $a=3$, the simulations produced data shown in Figs. \ref{overlay4} and \ref{overlay3}.  Fig. \ref{overlay4} overlays the simulation results and the mean field theoretical results shown in Fig. \ref{fig1} (using the same values of $\chi$).  One fitting parameter has been used for each $\epsilon_b$ to match the MFT and simulation results, namely the $\chi$ value that corresponds to each $\epsilon_b$.  $\chi$ should only depend on the interaction strength $\epsilon_b$ and the volume around each nanoparticle in which interactions occur, not on other details of the system.  Hence, the same $\chi$ fitting parameters were used in both Figs. \ref{overlay4} and \ref{overlay3}, demonstrating that for a given $\epsilon_b$, $\chi$ does not depend on $a$.

\begin{figure}[h]
\includegraphics[scale=0.5]{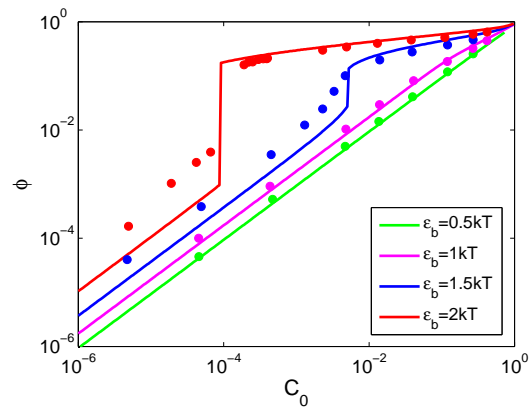}
\caption{$\phi$ vs. $C_0$ for four different binding strengths and grafting distance $a=4$.  Grand canonical simulation results (circles) and MFT results (lines) plotted on the same axes. The $\chi$ values for the analytical lines are the same as those of Fig. \ref{fig1} and correspond to the $\epsilon_b$ values shown in the legend.}
\label{phivsC0}
\end{figure}

Fig. \ref{phivsC0} shows the nanoparticle concentration in the polymer layer $\phi$ vs. the nanoparticle concentration in solution $C_0$.  This figure also shows qualitative agreement between MFT and simulation, including the presence of a phase transition for $\epsilon_b=2$ and $\chi=-13.5$.  The quantitative discrepancies between MFT and simulation are largest when the number of nanoparticles is small and the binding strength is high.  In this limit, there will be strong correlations between the locations of nanoparticles and monomers, and hence the mean-field picture of uniform density and random mixing begins to break down.

From the fitting of a $\chi$ value to each $\epsilon_b$, we found that $|\chi|$ increases approximately linearly with $\epsilon_b$.  This is a sensible result since the free energy of interaction, $a^2h\chi \phi \psi$ according to Eq.$\;$(\ref{Fkap-nup}), should in simulation be comparable to the number of interacting bead pairs times the strength of the interaction, which is on the order of $- \epsilon_b$.  The number of interacting pairs is roughly the number of nanoparticles $N_B=a^2h\phi$ times the number of monomer neighbors that each nanoparticle interacts with.  The number of monomer neighbors would be the volume around the ``tagged'' nanoparticle in which interactions occur, $V_{int}$, times the density of monomers $\psi$.  This means that the free energy of interaction in simulation should be roughly $-a^2h \epsilon_b V_{int} \phi \psi$.  Equating the mean field and simulation energies suggests $\chi \sim - V_{int} \epsilon_b$.  Since the effective range of the LJ force is of the order of $2b$, one expects $V_{int}$ to be of the order of $2^3$ in units of $b^3$.  A linear fit for $\chi(\epsilon_b)$ yields $\chi = -8.45\epsilon_b +4$.  Of course, this argument is overly simplistic, so only rough agreement can be expected.  One could also compare $\chi(\epsilon_b)$ to the second virial coefficient for the interaction.  Although the dependence of the second virial coefficient on $\epsilon_b$ is nonlinear, the range of $\epsilon_b$ considered here is small enough that the simplified linear approximation will suffice.
\subsection{Phase Transition}
	For sufficiently high  interaction strength, mean field theory predicts a discontinuity in the monomer and nanoparticle densities at the first-order transition.  This prediction is consistent with the simulation results of Figs. \ref{overlay4} and \ref{overlay3}.

	In order to probe the region very close to the phase transition, it proved convenient to perform simulations with a fixed number of nanoparticles.  In addition to speeding up simulations and providing easier access to situations in which very similar values of $C_0$ may yield very different equilibrium states, these simulations may be relevant to experimental setups in which the system size is small and the number of available nanoparticles in limited.  We refer to these as ``canonical ensemble simulations.''  In this type of simulation,  nanoparticles are still free to partition between the polymer layer and solution above, but they cannot enter or leave the simulation box.  In this case, instead of being fixed initially by the boundary conditions, the concentration of free nanoparticles above the brush $C_0$ was determined after equilibrium was established by tracking the mean number of nanoparticles which partitioned into the solution.

	Fixing the number of nanoparticles enabled us to observe the layer morphologies in the transition regime in which the polymer layer could not (according to mean field theory) attain either the extended phase or the collapsed phase globally.  Analogies to simpler systems would suggest that this situation could result in a coexistence region characterized by phase separation.

\begin{figure}[h]
\includegraphics[scale=0.5]{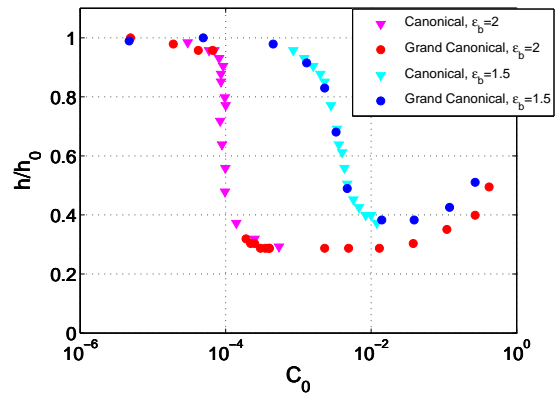}
\caption{Grand canonical simulation results from Fig. \ref{overlay4} overlaid with simulation results obtained by a canonical ensemble simulation method in which $C_0$ is not constrained by the boundary conditions.  Both sets of data are for $a=4$.}
\label{GCvsC}
\end{figure}

	As one can see in Fig. \ref{GCvsC}, canonical simulations and grand canonical simulations agree at concentrations for which data in available in both ensembles.  In addition, for $\epsilon_b=2$ the canonical simulations attain the same value of $C_0$ for a range of different values of $N_B$ and $h$.  This is a signature of a vertical drop in $h$ vs $C_0$, leading us to conclude that there is a phase transition for $\epsilon_b=2$.  The drop occurs at $C_0=C_0^*$, which is estimated to be between $8 \times 10^{-5}$ and $1 \times 10^{-4}$.  On the other hand, the curve for $\epsilon_b=1.5$ lacks such a vertical drop, leading us to conclude that $\epsilon_b=1.5$ is not large enough for a phase transition to occur.  This is consistent with the picture of a first order phase transition existing for sufficiently strong binding, but not for weak binding, as suggested by the mean field considerations.

Although most simulations were performed on a system of 16 polymer chains ($4 \times 4$), the canonical simulations were also performed with 36 chains ($6 \times 6$) in order to better visualize the system and test for finite size effects.  The 36 chain simulations (not shown) also show a vertical drop in $h$ vs. $C_0$ with the same $C_0^*$ as the 16 chain simulations (to within the uncertainty range indicated above).

\begin{figure}[h]
\subfloat{\includegraphics[height=2in]{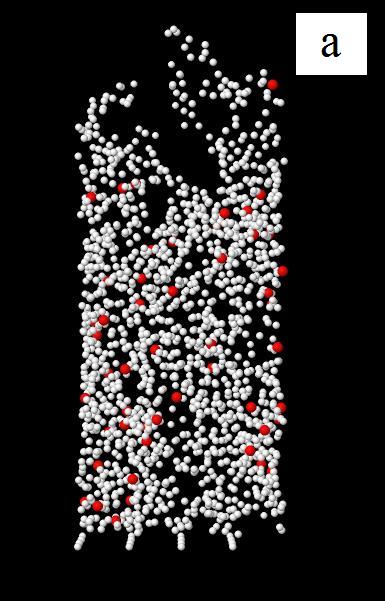}}
\hspace{8pt}
\subfloat{\includegraphics[height=2in]{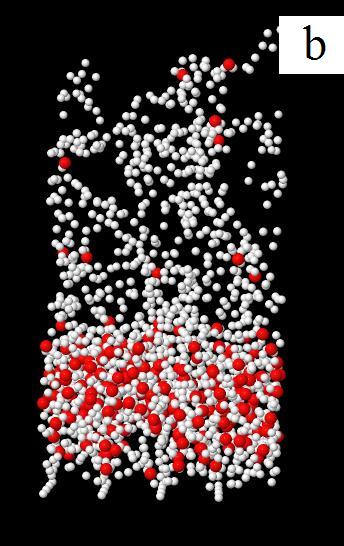}}\\
\subfloat{\includegraphics[height=2in]{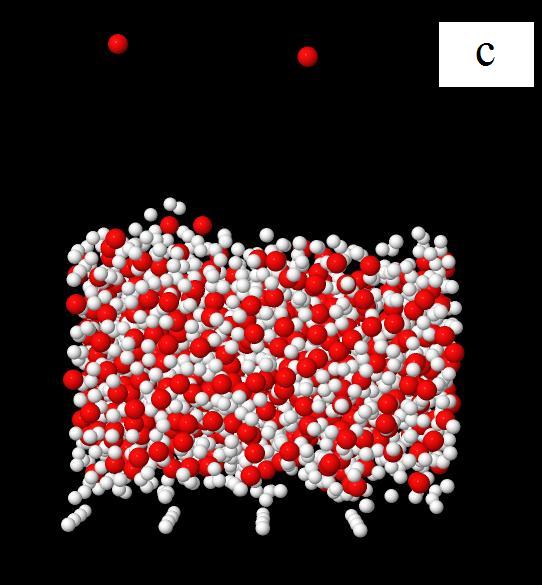}}
\hspace{8pt}
\subfloat{\includegraphics[height=2in]{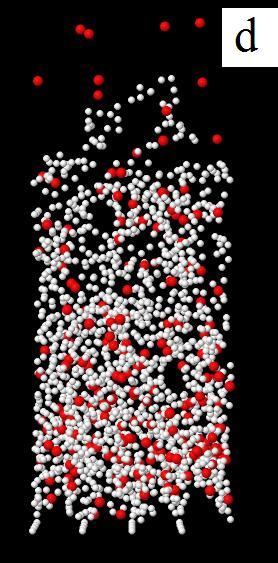}}
\caption{(a)-(c): Snapshots from a simulation with $\epsilon_b=2$ and $a=4$ at various values of $C_0$. (a): Just below the phase transition ($C_0=6.9\times10^{-5}$) the brush is extended and appears homogeneous. (b): At an intermediate brush height ($h/h_0 = 0.81$, $C_0=1.0 \times 10^{-4}$)  the brush appears to display spatial inhomogeneities. (c): Just above the phase transition ($C_0=2.4\times10^{-4}$) the brush is collapsed and appears homogeneous. (d): A snapshot from a simulation with $\epsilon_b=1.5$ and $a=4$ at an intermediate brush height ($h/h_0=0.84$) lacks the obvious spatial inhomogeneities of (b).}
\label{snapshots}
\end{figure}

Snapshots taken from equilibrated simulations can be used to show morphology changes in the phase transition region.  Fig. \ref{snapshots} shows snapshots taken from four of the simulations considered in Fig. \ref{GCvsC}.  For a concentration just below $C_0^*$ (Fig. \ref{snapshots}a), the polymer layer appears to be homogeneous and extended, while for a concentration just above $C_0^*$ (Fig. \ref{snapshots}c), the layer appears to be homogeneous and collapsed.  The canonical ensemble data allows us to access an intermediate case.  Fig. \ref{snapshots}b shows a brush with $\epsilon_b=2$ which has a number of bound nanoparticles such that it has an intermediate height ($h/h_0=0.81$).  This snapshot does not show the system to be in a single, homogenous phase, but rather shows a region of high density at low $z$ with a region of lower density above it.  In contrast, Fig. \ref{snapshots}d shows a snapshot of another polymer layer with a lower binding strength ($\epsilon_b=1.5$) and a number of nanoparticles such that $h/h_0=0.84$.  In this case, the polymer layer appears to remain in a homogeneous phase despite its intermediate height.

\subsection{Density Profiles}\label{Profiles}
\begin{figure}[h]
\includegraphics[scale=0.5]{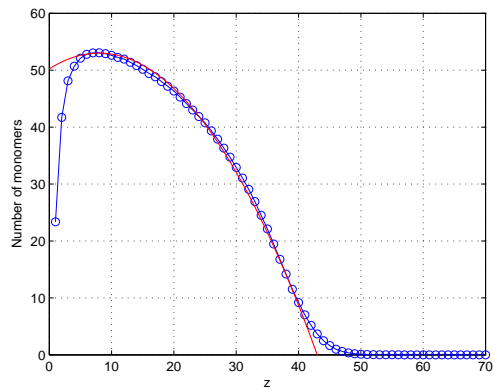}
\caption{A time averaged histogram of the monomers' height above the grafting surface for parameter values for which few nanoparticles are bound to the brush.  The monomer density profile is nearly parabolic as expected for a brush with no nanoparticles.  The red line shows a parabola to guide the eye.  This figure was generated for $a=4$, $\epsilon_b =2$, $C_0 = 4.2\times 10^{-5}$.}
\label{fig_par}
\end{figure}

When the nanoparticle concentration is small enough that they constitute only a small perturbation on the standard polymer brush, the monomer density profile is approximately parabolic as seen in Fig. \ref{fig_par}.  The height of the brush is approximately the intercept of the fit parabolia as shown in Fig. \ref{fig_par}.  Although our MFT takes the monomer density to be constant throughout the polymer layer, the parabolic density profile obtained in simulations matches the expected behavior of a plane-grafted brush \cite{MilnerWittenCates, Milner1991}, including the presence of a depletion region near the wall and a ``foot'' at high $z$ in simulations \cite{Murat1989, Milner1990}.

\begin{figure}[h]
\includegraphics[scale=0.5]{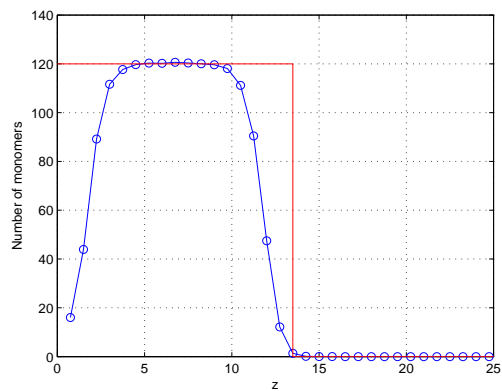}
\caption{A time averaged histogram of the monomers' height above the grafting surface for parameter values for which many nanoparticles are bound to the brush.  The red line shows a step function which drops to zero at the measured value of h.  The monomer density profile is close to a step function because monomers and nanoparticles are tightly packed as in a solid.  This figure was generated for $a=4$, $\epsilon_b =2$, $C_0 = 2.3\times 10^{-3}$.}
\label{step_histogram}
\end{figure}

Under certain conditions, the brush collapses, and space is almost completely filled with particles.  As a result, the monomer density profile becomes approximately a step function as seen in Fig. \ref{step_histogram}.  Of course in simulations the step is not infinitely sharp, but instead includes a transition region of intermediate monomer concentration.  We include this intermediate region as part of the brush.


\begin{figure}[h]
\includegraphics[scale=0.5]{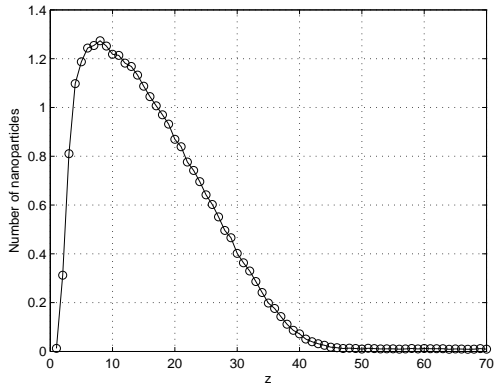}
\caption{A time averaged histogram of the nanoparticles' height above the grafting surface for the same parameters as Fig. \ref{fig_par}.}
\label{fig_par2}
\end{figure}

\begin{figure}[h]
\includegraphics[scale=0.5]{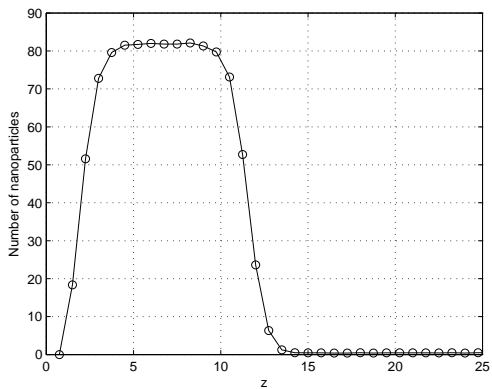}
\caption{A time averaged histogram of the nanoparticles' height above the grafting surface for the same parameters as Fig. \ref{step_histogram}.}
\label{step_histogram2}
\end{figure}

Typical nanoparticle density profiles corresponding to Figs.  \ref{fig_par} and \ref{step_histogram} can be seen in Figs. \ref{fig_par2} and \ref{step_histogram2}.  In both cases, the nanoparticle density profile for $z<h$ qualitatively tracks the monomer density profile because nanoparticles in the polymer layer are expected to be bound to one or more monomers.

\section{Summary and Discussion}
In this paper, we considered the case of grafted polymers in equilibrium with nanoparticles that are attracted to the polymers.  The mean field theory of Sec. \ref{MFT} and the Brownian dynamics simulations of Sec. \ref{SimResults} are in qualitative agreement that, at certain interaction strengths and nanoparticle concentrations, the grafted polymer layer is expected to attain a more compact state upon the addition of binding nanoparticles.  Hence, our results are in qualitative agreement with the biologically-motived experiments of Ref.$\;$\cite{Lim2007}.

The results of mean field theory and Brownian dynamics simulations not only show that the polymer layer is expected to attain a more compact state, they also reveal the possibility of a discontinuous collapse of a polymer brush in the presence of attracting nanoparticles as a function of nanoparticle concentration in the abutting solution.  For sufficiently weak binding the collapse of the brush is smooth, while for sufficiently strong binding, the collapse occurs via a first-order phase transition from an extended state to a collapsed state.

These observations may be useful in the field of polymer science.  Control of polymer morphology in the presence of additives not only plays a role in the development of tougher polymeric materials (via the use of antiplasticizers, for example), but polymeric materials (sometimes decorated with nanoparticles of various kinds, e.g. \cite{SolarCells,VirusMemory,NanoparticleMemory}) are also becoming increasingly useful for applications such as organic electronics.

In addition, the insights gained in this study inform future analysis of the behavior of grafted biopolymers, such as those in the Nuclear Pore Complex (NPC) in which binding proteins play the role of nanoparticles \cite{Schulten2005, Stewart2005}.  Of course, in physically realistic situations, the theory and simulations presented here may be overly simplistic.  In the relevant biophysical system of the nuclear pore complex, which is involved in the transport of proteins and nucleic acids into and out of the cell nucleus, the ``nanoparticles" (which correspond to binding proteins) are significantly larger than those of our simulations \cite{Blobel1999}, and they possess discrete binding sites rather than the isotropic binding potential used here.  Complications like these will be addressed in future work.  In spite of its simplicity, the model presented here appears to share qualitative similarities with the NPC analogs, such as the collapse of nucleoporins upon binding to karyopherin transport proteins which was observed by Lim {\it et al.} \cite{Lim2007}.

\section{Acknowledgments}
We thank Roderick Lim, Jaclyn Tetenbaum-Novatt, Siegfried Musser, Tijana Jovanovic-Talisman, Yitzhak Rabin, Michael Rexach, Michael Rout, and Paul Welch for helpful discussions.  AZ acknowledges the support from NSERC and DOE.
RDC acknowledges financial support from NSF grant CHE-0750332.
\bibliography{collapsepaper}
\end{document}